\documentclass[aps,prc,twocolumn,superscriptaddress,showpacs]{revtex4-1}

\usepackage{soul}
\usepackage{color}
\usepackage{amssymb}
\usepackage{graphicx}
\usepackage[normalem]{ulem}

\begin{document}
\title{Using Neutron Star Observations to Determine
Crust Thicknesses,\\Moments of Inertia, and Tidal Deformabilities}

\author{A.~W. Steiner}
\affiliation{Institute for Nuclear Theory, University of 
Washington, Seattle, Washington 98195, USA}
\affiliation{Department of Physics and Astronomy, University of
Tennessee, Knoxville, Tennessee 37996, USA}
\affiliation{Physics Division, Oak Ridge National Laboratory, Oak
Ridge, Tennessee 37831, USA}

\author{S. Gandolfi}
\affiliation{Theoretical Division, Los Alamos National Laboratory, Los
Alamos, New Mexico 87545, USA}

\author{F.~J. Fattoyev}
\affiliation{Department of Physics and Astronomy, 
  Texas A\&M University-Commerce, Commerce, Texas 75429, USA}
\affiliation{Department of Physics and Nuclear Theory Center,
Indiana University, Bloomington, Indiana 47405, USA}

\author{W.~G. Newton}
\affiliation{Department of Physics and Astronomy, 
Texas A\&M University-Commerce, Commerce, Texas 75429, USA}

\begin{abstract}
We perform a systematic assessment of models for the equation of state
(EOS) of dense matter in the context of recent neutron star mass and
radius measurements to obtain a broad picture of the structure of
neutron stars. We demonstrate that currently available neutron star
mass and radius measurements provide strong constraints on moments of
inertia, tidal deformabilities, and crust thicknesses. A measurement
of the moment of inertia of PSR J0737-3039A with 10\% error, without
any other information from observations, will constrain the EOS over a
range of densities to within 50\%$-$60\%. We find tidal deformabilities
between 0.6 and $6\times 10^{36}$ g cm$^{2}$ s$^{2}$ (to 95\%
confidence) for $M=1.4~\mathrm{M}_{\odot}$, and any measurement which
constrains this range will provide an important constraint on dense
matter. The crustal fraction of the moment of inertia can be as large
as 10\% for $M=1.4~\mathrm{M}_{\odot}$ permitting crusts to have a
large enough moment of inertia reservoir to explain glitches in the
Vela pulsar even with a large amount of superfluid entrainment.
Finally, due to the uncertainty in the equation of state, there is at
least a 40\% variation in the thickness of the crust for a fixed mass
and radius, which implies that future simulations of the cooling of a
neutron star crust which has been heated by accretion will need to
take this variation into account.
\end{abstract}

\pacs{26.60.-c, 21.65.Cd, 26.60.Kp, 97.60.Jd }
\preprint{INT-PUB-14-008} 
\preprint{LA-UR-14-21809}

\maketitle

\section{Introduction}

Recent neutron star (NS) mass and radius observations have provided
new constraints on the neutron star mass-radius curve and on the
equation of state (EOS) of dense matter~\cite{Lattimer01}. The EOS, in
turn, is a fundamental property of quantum chromodynamics, which
probes cold and dense matter which is otherwise difficult to access in
experiment.

In the near future, mass and radius observations may be complemented
by other constraints on NS structure. Although thousands of pulsars have
been observed, there is only one binary system where both NSs are
radio-active pulsars, PSR J0737-3039. The ability to observe
pulsations from both NSs and the extreme nature of the
system~\cite{Burgay03,Kramer06}, enables a potential measurement of
the moment of inertia of one of the neutron
stars~\cite{Damour88}. Also, the Laser Interferometer
Gravitational-Wave Observatory is expected to measure the
gravitational wave signal from a NS merger within the near
future~\cite{Abadie10}, and a sufficiently large signal to noise
observation will enable a measurement of the neutron star tidal
deformability~\cite{Flanagan08,Hinderer10,Damour12,delPozzo13,
  Read13,Lackey14} (denoted by $\lambda$ and sometimes also called
``tidal polarizability''). It turns out these two types of new
observations are intimately related: the moment of inertia of a NS is
strongly correlated with its tidal
deformability~\cite{Yagi13a,Yagi13b}.

NSs can accrete matter from main-sequence companions which results in
the emission of X-rays and the heating of the NS crust. If the
accretion stops (referred to as ``quiescence'' since the X-rays from
accretion subside), then the cooling NS crust can be directly
observed~\cite{Rutledge02}. The timescale for this cooling is
proportional to the square of the NS crust
thickness~\cite{Lattimer94}, and thus the crust thickness is important
for determining the properties of the crust from observations of crust
cooling~\cite{Lewin06,Shternin07,Brown09,Page13}.

Another potential constraint of NS structure comes from pulsar
glitches. Previous papers~\cite{Link99,Andersson12,Chamel13} have shown
that, if NS crusts are believed to be the location of the angular
momentum reservoir which contributes to the glitch spin up, then a
significant fraction of the NS's moment of inertia must lie in the
superfluid component of the crust. Thus glitches are sensitive to the
crustal fraction of the moment of inertia, denoted $\Delta I/I$.

Finally, many of these quantities are (at
least weakly) correlated with the nuclear symmetry
energy~\cite{Steiner05,Tsang12,Li14}. The nuclear symmetry energy is the
difference between the energy per baryon of neutron matter and that of
nuclear matter (we ignore quartic terms, see Ref.~\cite{Steiner06}).
We denote the symmetry energy $S(n_B)$, where $n_B$ is the baryon
number density, and $S\equiv S(n_0)$, where $n_0$ is the nuclear
saturation density. The quantity $3 n_0 S^{\prime}(n_0)$ is denoted as
$L$. The value of $L$ determines the pressure of neutron-rich matter
at the saturation density. The pressure of neutron-rich matter, in
turn, is related to all of the above NS structure quantities given
above. 

For the first time, we use existing NS mass and radius observations to
predict the expected ranges of NS properties such as moments of
inertia, tidal polarizabilities and crustal thicknesses which are
measurable by a diverse range of ongoing observational programs. We
generate these expected ranges based on Monte Carlo simulations using
parameterizations which explore the full variation which is possible
given current uncertainties in the nature of dense matter. Our EOS
models are based on recent progress in the microscopic calculation of
neutron-rich matter near the nuclear saturation density. At high
densities, we assume no additional correlation with matter at lower
densities, and use models which allow for strong phase transitions.
Our method is in contrast to several previous papers which have
computed theoretical predictions of moments of inertia, crust
thicknesses, and tidal polarizabilities for smaller samples of
representative
EOSs~\cite{Kalogera99,Lattimer01,Morrison04,Bejger05,Steiner05,Lattimer05,
  Lattimer07,Fattoyev10,Fattoyev13,Lattimer13}. Future observations
will consitute direct tests of the theoretical framework we use and of
the systematics of current mass-radius observations.

\section{Method}

For the first observational data set, we use (i) the five mass and
radius measurements from photospheric radius expansion (PRE) bursts in
Ref.~\cite{Ozel10,Steiner10,Steiner13,Lattimer14a} (by assuming that
the photosphere is extended at ``touchdown'' as justified in
Ref.~\cite{Steiner10}) and (ii) the five radius measurements from
quiescent low-mass X-ray binaries in
Refs.~\cite{Guillot13,Lattimer14b} by taking the hydrogen column
densities and distances from the Harris catalog~\cite{Harris96} and
allowing for either hydrogen or helium atmospheres (this is the choice
from Ref.~\cite{Lattimer14b} with the largest value for the evidence
integral). The second data set additionally assumes a hypothetical
10\% measurement of $I=(70{\pm}7)~\mathrm{M}_{\odot}~\mathrm{km}^2$
for a 1.4-$\mathrm{M}_{\odot}$ NS. The centroid of this value is near
that predicted by the mass-radius data (to allow easier comparison).
The third type of data set includes only the $I$ measurement (either
$70\pm 7$, $80\pm 8$, or $90\pm 9$ $\mathrm{M}_{\odot}~\mathrm{km}^2$)
and no constraint from other mass and radius observations. Larger
values of $I$ would be implied by the radius constraint from long PRE
bursts as suggested in Ref.~\cite{Suleimanov11}.

There is a one-to-one correspondence between the NS mass-radius curve
and the pressure as a function of energy density
$P(\varepsilon)$. We ensure that all EOSs are causal
($dP/d\varepsilon<1$), hydrodynamically stable $dP/d\varepsilon>0$,
and that all mass-radius curves produce a 2-$\mathrm{M}_{\odot}$ NS in line
with the recent mass measurements in
Refs.~\cite{Demorest10,Antoniadis13}. Strange quark matter is assumed
not to be absolutely stable, so we consider only hybrid NSs where
deconfined quark matter is surrounded by a hadronic crust and leave
the consideration of strange quark stars to future work. Moments of
inertia are computed using the slow rotation (Hartle-Thorne)
approximation~\cite{Hartle67, Hartle68}, and we use the correlation in
Refs.~\cite{Yagi13a,Yagi13b} to compute the tidal deformabilities. We
have independently checked this correlation based on the expressions
in Refs.~\cite{Postnikov10,Hinderer10} and find that the correlation
generally holds to within about 1\% (since we are ignoring strange
quark stars), although slightly larger variations can be generated with
strong phase transitions just above the nuclear saturation density.
However, such configurations, although not ruled out by the observational
data, are finely-tuned and relatively improbable, and thus the results
from the correlation are sufficient for our purposes.

There has been significant recent
progress~\cite{Gandolfi12,Hebeler13,Gezerlis13} on computing the EOS
of neutron rich matter from using realistic nuclear forces, both
quantum Monte Carlo and using chiral effective theory interactions in
many-body perturbation theory. We assume that the binding energy of
nuclear matter is $-16$ MeV, the saturation density is 0.16 fm$^{-3}$
(typical values from Ref.~\cite{Kortelainen10}), and we choose limits
for the incompressibility of $220~\mathrm{MeV}<K<260~\mathrm{MeV}$
from Refs.~\cite{Shlomo06,Piekarewicz10}. Two different EOSs are
employed near the nuclear saturation density. The first is the quantum
Monte Carlo model from Ref.~\cite{Gandolfi12}, and we refer to this
model as ``Gandolfi-Carlson-Reddy (GCR)''. The limits
$12.5~\mathrm{MeV}<a<13.5~\mathrm{MeV}$ and $0.47<\alpha<0.53$ are
increased slightly from Ref.~\cite{Steiner12} to ensure that we
include all possible models from Ref.~\cite{Gandolfi12}. These two
parameters principally parametrize the two-nucleon part of the
interaction. Additionally, we reparametrize $b$ and $\beta$,
parameters which control the three-nucleon interaction, in terms $S$
and $L$. We limit $S$ to between 29.5 and 36.1 MeV to be
consistent with the second model described below. We limit $L$ to be
between 30 and 70 MeV which covers the range of $L$ from
Refs.~\cite{Gandolfi12,Gandolfi14}. We note that the GCR model is
essentially a sum of two polytropes, with coefficients and exponents
that were constrained by experiment.

For the second model, we use the parametrization from
Ref.~\cite{Hebeler13} [``Hebeler-Lattimer-Pethick-Schwenk (HLPS)''],
and the results on neutron matter from Ref.~\cite{Tews13} from an
interaction based on chiral effective theory. At each point, we fix
$\alpha, \gamma, \eta$, the three parameters which control nuclear
saturation, to fix the saturation density, the binding energy, and the
incompressbility. Note that this $\alpha$ is distinct from the
parameter with the same name in the GCR model. The remaining two
parameters $\alpha_L$ and $\eta_L$, which control the properties of
neutron matter, are again reparameterized in terms of $S$ and $L$. The
range of $S$ from Fig.~1 of Ref.~\cite{Tews13} is between 29.5 and
36.1 MeV and the range of $L$ is between $44.0$ and $65.0$ MeV.

Both nuclear masses and theoretical models imply a correlation between
$S$ and $L$, thus we additionally restrict parameters to lie between
$(9.17~S-266~\mathrm{MeV}) < L < (14.3~S-379~\mathrm{MeV})$, which
encloses the constraints from nuclear masses~\cite{Kortelainen10},
quantum Monte Carlo~\cite{Gandolfi12}, chiral
interactions~\cite{Tews13}, and isobaric analog
states~\cite{Danielewicz14} as summarized in Ref.~\cite{Lattimer14a}.
The GCR and HLPS models are used only up to the nuclear saturation
density as they may not be valid if a phase transition is present.
By increasing the density up to which we use these models would improve
our constraints on the EOS but does not change the qualitative
results. It is particularly critical that we assume no correlation
between the EOS near the saturation density and the EOS at higher
densities, except for the constraint that $P(\varepsilon)$ is a
continuous and monotonically increasing function.
 
For the NS crust, we use the tabulated crust EOSs based on the work in
Ref.~\cite{Newton11}. The advantage of this crust EOS, relative to the
older work in Ref.~\cite{Baym71}, is that we can employ the same
values of $S$ and $L$ which we use in the EOS of neutron-rich matter
at the saturation density. A two-dimensional grid of crust EOSs with
varying values of $S$ and $L$ were computed and this grid was
interpolated to generate the crust for general values of $S$ and $L$
in our simulations. For the transition between the crust and the core,
we use the correlation between $n_t$ and $S$ and $L$,
\begin{equation}
n_t = S_{30} \left(0.1327 - 0.0898 L_{70} + 
0.0228 L_{70}^2\right)~\mathrm{fm}^{-3}
\label{eq:ntcorr}
\end{equation}
where $L_{70}$ is $L$ in units of 70 MeV and $S_{30}$ is $S$ in units of
30 MeV. For our ranges of $S$ and $L$, this correlation gives
transition densities between 0.06 fm$^{-3}$ and 0.1 fm$^{-3}$,
consistent with those obtained in Ref.~\cite{Oyamatsu07}. We use this
transition density from Eq.~\ref{eq:ntcorr} to compute the transition
pressure, and find values between 0.30 and 0.82 MeV/fm$^{3}$, slightly
larger than the range 0.25$-$0.65 MeV/fm$^{3}$ found in
Ref.~\cite{Lattimer01}.

For the EOS above the saturation density, we either use a set of three
piecewise polytropes (only five parameters since the transition to the
first polytrope is already fixed by the EOS at the saturation density)
referred to as ``Model A'' in Ref.~\cite{Steiner13}. Alternatively, we
use a set of four line segments in the $(\varepsilon,P)$ plane,
``Model C''~\cite{Steiner13}. This latter model is useful because it
provides an alternative model which tends to favor stronger phase
transitions in the core. We do not employ Model B or Model D from
Ref.~\cite{Steiner13} because they do not typically provide
significantly different results from Model A at the current level of
accuracy. The choice of either GCR or HLPS near saturation density and
either Model A or Model C at high densities gives a total of four EOS
models to use with our three data sets.

For each of the above data sets and EOS models, we perform a Markov
chain Monte Carlo simulation as first outlined in
Ref.~\cite{Steiner10}. To obtain our final results for a fixed data
set we choose the smallest range which encloses all of the EOS models,
as performed in Ref.~\cite{Steiner13}. This procedure is a relatively
simple version of a fully hierarchical Bayesian analysis which is
currently too computationally expensive. 

\begin{table*}
\begin{tabular}{c|cc|cc|cc}
& \multicolumn{2}{c}{GCR, Mod. A, $M$\&$R$ data} &
\multicolumn{2}{c}{GCR, Mod C, $M$\&$R$ data} &
\multicolumn{2}{c}{GCR, Mod. A, $I=90$} \\
Quantity (unit) & 95\% lower & 95\% upper &
95\% lower & 95\% upper &
95\% lower & 95\% upper \\
\hline
$P(\varepsilon=300)~(\mathrm{MeV}/\mathrm{fm}^3)$ & 
9.318 & 22.86 & 2.253 & 15.83 & 20.39 & 65.70 \\
$P(\varepsilon=450)~(\mathrm{MeV}/\mathrm{fm}^3)$ & 
33.31 & 71.02 & 25.88 & 68.09 & 60.53 & 132.1 \\
$P(\varepsilon=600)~(\mathrm{MeV}/\mathrm{fm}^3)$ & 
90.98 & 160.8 & 76.56 & 216.3 & 104.6 & 220.3 \\
$P(\varepsilon=1000)~(\mathrm{MeV}/\mathrm{fm}^3)$ & 
281.0 & 413.0 & 260.9 & 558.1 & 260.4 & 461.2 \\
$L~(\mathrm{MeV})$ & 
30.53 & 65.79 & 30.53 & 68.41 & 35.64 & 69.66 \\
$n_t~(\mathrm{fm}^{-3})$ & 
0.07142 & 0.09903 & 0.07168 & 0.1021 & 0.07061 & 0.09766 \\
$P_t~(\mathrm{MeV}/\mathrm{fm}^{3})$ & 
0.3125 & 0.8163 & 0.3149 & 0.7968 & 0.3119 & 0.8006 \\
$\Delta R(n_t=0.06,M=1.4)$ & 
0.6904 & 1.037 & 0.5949 & 0.9282 & 0.9550 & 1.488 \\
$\Delta R(n_t=0.08,M=1.4)$ & 
0.7455 & 1.206 & 0.6483 & 1.088 & 1.038 & 1.729 \\
$\Delta R(n_t=0.10,M=1.4)$ & 
0.8025 & 1.256 & 0.7073 & 1.136 & 1.136 & 1.863 \\
$\Delta R(n_t=0.08,M=1.0)$ & 
1.164 & 1.881 & 0.9457 & 1.651 & 1.539 & 2.383 \\
$\Delta R(n_t=0.08,M=2.0)$ & 
0.3411 & 0.6475 & 0.3252 & 0.6401 & 0.5420 & 1.154 \\
$R(M=1.4)~(\mathrm{km})$ & 
10.79 & 12.44 & 10.22 & 11.87 & 12.39 & 14.47 \\
$R(M=1.7)~(\mathrm{km})$ & 
10.74 & 12.40 & 10.31 & 11.95 & 12.36 & 14.82 \\
$R(M=2.0)~(\mathrm{km})$ & 
10.16 & 12.25 & 10.10 & 12.01 & 11.96 & 15.13 \\
$R_{\mathrm{max}}~(\mathrm{km})$ & 
9.812 & 11.57 & 9.792 & 11.81 & 11.15 & 14.49 \\
$n_{B,\mathrm{max}}~(\mathrm{fm}^{-3})$ & 
0.8770 & 1.234 & 0.7642 & 1.235 & 0.6059 & 0.9794 \\
${\varepsilon}_{\mathrm{max}}~(\mathrm{MeV}/\mathrm{fm}^{3})$ & 
1055 & 1597 & 578.4 & 1612 & 624.7 & 1236 \\
$I(M=1.4)~(\mathrm{M}_{\odot}~\mathrm{km}^2)$ & 
60.62 & 77.06 & 56.25 & 69.87 & (fixed) & \\
$I(M=1.7)~(\mathrm{M}_{\odot}~\mathrm{km}^2)$ & 
80.11 & 101.5 & 77.02 & 95.49 & 99.06 & 141.5 \\
$I(M=2.0)~(\mathrm{M}_{\odot}~\mathrm{km}^2)$ & 
94.49 & 126.7 & 97.59 & 125.0 & 119.3 & 184.4 \\
$(\Delta I/I)(n_t=0.08,M=1.4)$ & 
0.02045 & 0.06084 & 0.01555 & 0.04723 & 0.03895 & 0.1033 \\
$(\Delta I/I)(n_t=0.08,M=1.7)$ & 
0.01284 & 0.03703 & 0.01033 & 0.03211 & 0.02346 & 0.07233 \\
$(\Delta I/I)(n_t=0.08,M=2.0)$ & 
0.006949 & 0.02317 & 0.006129 & 0.02231 & 0.01377 & 0.05424 \\
$\lambda(M=1.4)$ & 
1.000 & 2.606 & 0.7306 & 1.811 & 1.945 & 5.904 \\
$\lambda(M=1.7)$ & 
0.6596 & 2.067 & 0.5258 & 1.527 & 1.716 & 7.505 \\
$\lambda(M=2.0)$ & 
0.2039 & 1.276 & 0.2635 & 1.211 & 0.6898 & 6.865 \\
\hline
\end{tabular}
\caption{Predictions for the 95\% confidence limits. The second and
  third columns show results for GCR, Model A, and the mass and radius
  data set. Within this particular model, the EOS is constrained to
  within about a factor of 2, and neutron star radii are constrained
  to around 1.5 kilometers. Results for the HLPS model are very
  similar. Crust thicknesses $\Delta R$ are given in $\mathrm{km}$ and
  tidal deformabilities $\lambda$ are given in units of
  $10^{36}~\mathrm{g}~\mathrm{cm}^2~\mathrm{s}^2$. The fourth and
  fifth column show results for Model C instead of Model A. Finally,
  the sixth and seventh columns use only a measurement of
  $I=90\pm9~\mathrm{M}_{\odot}~\mathrm{km}^2$ for a
  1.4-$\mathrm{M}_{\odot}$ neutron star as a data set. This table
  summarizes the results for 3 of the 16 combinations of models and
  data sets used in this paper. }
\label{tab:GCRAMR}
\end{table*}

\section{Results} 

Results that use GCR for matter near saturation density, Model A for
higher densities, and that use the mass-radius data described above are
summarized in the second and third columns of Table~\ref{tab:GCRAMR}.
For this particular model and data set, the results on non-rotating
stars are very similar to those obtained
previously~\cite{Steiner12,Steiner13}. The moment of inertia ranges
between 60 and 130 $\mathrm{M}_{\odot}~\mathrm{km}^2$, with even
smaller values for lower mass neutron stars. The fraction of the
moment of inertia which lies in the neutron star crust is small,
between 2\% and 6\% for 1.4-$\mathrm{M}_{\odot}$ neutron stars. The
tidal deformability ranges between 0.2 and 2.6$\times 10^{36}$ g
cm$^2$ s$^2$, which depends on mass. The HLPS model gives very similar
neutron star radii in comparison to GCR. The numbers in the table
outline the limits of probability distributions taken from the Monte
Carlo. No assumption is made about the shape of the probability
distribution, and the distributions can be significantly non-Gaussian.

A probability distribution for all of the relevant quantities can be
generated for any combination of EOS model and neutron star data set,
and 16 combinations are explored in this paper. The variations among
models and data sets are summarized in Table~\ref{tab:radii} which
gives in 95\% confidence limits for the radius of a
1.4-$\mathrm{M}_{\odot}$ neutron star. Model C gives smaller NS radii
than Model A, due to the possible presence of phase transitions. A
measurement of $I=70$ implies slightly larger radii than that implied
by the mass and radius data, which is the result of the fact that the
mass and radius data prefer slightly smaller moments of inertia (i.e.
columns four and five in Table~\ref{tab:GCRAMR}). Larger $I$
measurements imply larger radii as large as 14.7 km in some cases.
These large neutron star radii are due to EOSs in which the pressure
becomes significantly larger above the saturation density, but these
are only likely if there is some systematic uncertainty which
invalidates the mass and radius observations which were described
above.

\begin{table}
\begin{tabular}{c|cc}
& \multicolumn{2}{c}{Radius for $M={1.4}~\mathrm{M}_{\odot}$ (km)} \\
Model & 95\% lower limit & 95\% upper limit \\
\hline
GCR, Mod. A, $M$\&$R$ &
10.79 & 12.44 \\
GCR, Mod. C, $M$\&$R$ &
10.22 & 11.87 \\
HLPS, Mod. A, $M$\&$R$ &
10.82 & 12.42 \\
HLPS, Mod. C, $M$\&$R$ &
10.21 & 11.86 \\
GCR, Mod. A, $M$\&$R$+$I=70$ &
11.12 & 12.57 \\
GCR, Mod. C, $M$\&$R$+$I=70$ &
10.47 & 12.03 \\
HLPS, Mod. A, $M$\&$R$+$I=70$ &
11.13 & 12.49 \\
HLPS, Mod. C, $M$\&$R$+$I=70$ &
10.50 & 12.09 \\
GCR, Mod. A, $I=70$ &
11.66 & 13.67 \\
GCR, Mod. C, $I=70$ &
10.55 & 13.47 \\
HLPS, Mod. A, $I=70$ &
11.66 & 13.43 \\
HLPS, Mod. C, $I=70$ &
10.64 & 13.64 \\
GCR, Mod. A, $I=80$ &
12.04 & 14.14 \\
GCR, Mod. C, $I=80$ &
11.36 & 14.41 \\
GCR, Mod. A, $I=90$ &
12.39 & 14.47 \\
GCR, Mod. C, $I=90$ &
10.59 & 14.68 \\
\end{tabular}
\caption{Limits for the radius of a 1.4-$\mathrm{M}_{\odot}$ neutron
star for various models and data sets.}
\label{tab:radii}
\end{table}

To allow more complete comparison between Model A and Model C,
Table~\ref{tab:GCRAMR} tabulates the results for Model C for the same
EOS at lower densities (GCR) and the same data set. The strong phase
transitions implied by Model C lead to smaller pressures at low energy
densities and higher pressures at high energy densities. This result
originates in the constraints from observations (which require small
radii from low-mass neutron stars), and the constraint of a
2-$\mathrm{M}_{\odot}$ neutron star (which requires a higher pressure
at higher densities). Generally, radii, moments of inertia, and tidal
deformabilities are smaller. Values of $L$ are potentially a bit
larger in Model C; strong phase transitions permit a higher pressure
near the saturation density because they make up for them with a lower
pressure at higher densities. Constraints on $S$ are not reported,
because neutron star observations do not currently constrain $S$. The
interplay between density regimes is difficult to observe in other
studies which presume some sort of correlation between the nature of
matter at saturation density and at higher densities.

\begin{figure}
\includegraphics[width=3.4in]{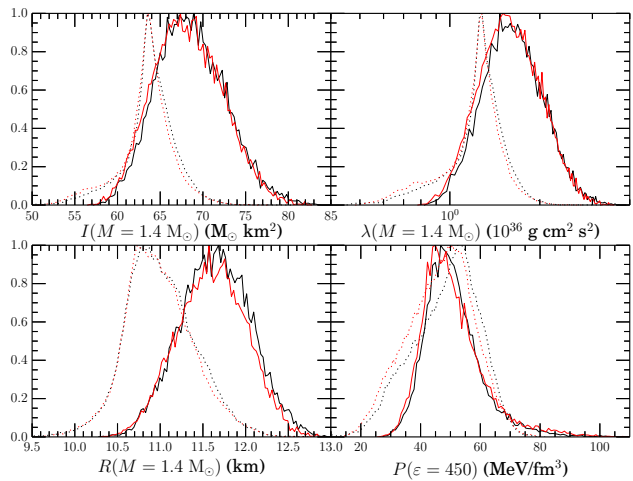}
\caption{Probability distributions for the moment of inertia of a
  1.4-$\mathrm{M}_{\odot}$ NS (upper left), tidal deformability of a
  1.4-$\mathrm{M}_{\odot}$ NS (upper right), radius of a
  1.4-$\mathrm{M}_{\odot}$ NS (lower left) and pressure at about three
  times the saturation density (lower right) given the set of mass and
  radius observations. Black lines are for GCR and red lines are for
  HLPS. Solid lines are for Model A and dotted lines are for Model C.
  Each of the distributions was separately normalized to have a
  maximum at 1.\label{fig:hist}}
\end{figure}

Mass and radius observations suggest probability distributions for the
moment of inertia of a 1.4-$\mathrm{M}_{\odot}$ neutron star as given
in the upper left panel of Fig.~\ref{fig:hist}. It is clear that Model
C, which favors stronger phase transitions gives slightly smaller
values for $I(M=1.4~\mathrm{M}_{\odot})$ as expected. Among the four
models which are plotted, the smallest 68\% lower limit is 61.4
M$_{\odot}$ km$^{2}$ and the largest 68\% lower limit is 72.7
M$_{\odot}$ km$^{2}$, i.e., a variation of less than 20\%. The
corresponding range for the radius of a 1.4 M$_{\odot}$ NS is
10.6$-$12.1 km, comparable to results obtained
previously~\cite{Steiner10,Steiner12,Steiner13,Lattimer14b}. The
relative size of the constraint on the pressure at an energy density
of 450 MeV/fm$^{3}$ from the $M-R$ data is 53\%. On the other hand,
since the systematic uncertainties of currently available mass and
radius observations may be larger than that from a future $I$
measurement, it is worth noting that a 10\% $I$ measurement alone
constrains the pressure to within 55\% at that same energy density and
to within 59\% at an energy density of 1000 MeV/fm$^{3}$.

Current mass and radius observations imply tidal deformabilities for a
1.4-$\mathrm{M}_{\odot}$ NS between (1.09 and 2.12)$\times 10^{36}$ g
cm$^{2}$ s$^{2}$ to 68\% confidence over all four EOS models. These
provide guidance on how sensitive GW observatories will likely need to
be to detect the tidal deformation in a double neutron star
merger~\cite{Hinderer10,Lackey14}. This result is a natural
consequence of rather small neutron star radii implied by the
quiescent low-mass X-ray binaries
(qLMXBs)~\cite{Guillot13,Lattimer14b}. Over all four EOS models none
of the 95\% confidence limits goes higher than 2.6 $\times 10^{36}$ g
cm$^{2}$ s$^{2}$.

On the other hand, if we assume that the systematic uncertainties
spoil mass and radius observations, tidal deformabilities can be
larger. The 95\% confidence limits of various quantities that assume a
measurement of $I=(90{\pm}9)~\mathrm{M}_{\odot}~\mathrm{km}^2$ are
summarized in the sixth and seventh columns of Table~\ref{tab:GCRAMR}.
This case shows tidal deformabilities can easily be as large as 7-8
$\times 10^{36}$ g cm$^{2}$ s$^{2}$, depend on mass. The most extreme
95\% limits for the tidal deformability of a 1.4-$\mathrm{M}_{\odot}$
over all 16 combinations of models and data are 0.64 and 6.1 $\times
10^{36}$ g cm$^{2}$ s$^{2}$.

\begin{figure}
\includegraphics[width=3.4in]{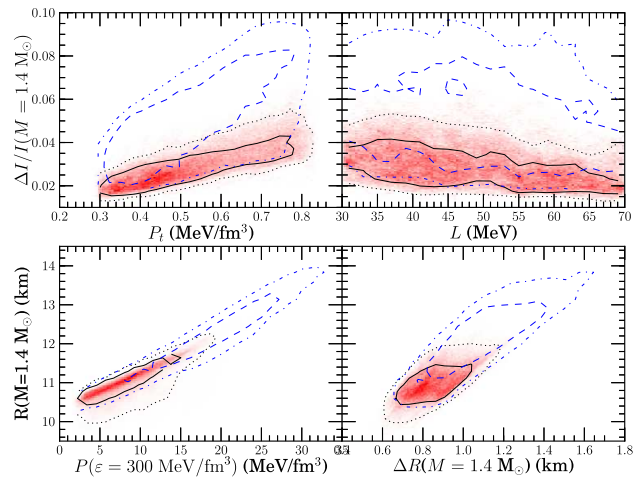}
\caption{Probability distributions for four pairs of quantities with
  EOS model GCR and Model C at high densities. In the upper-left
  panel, $P_t$ is the pressure at the core-crust transition. All
  other quantities are defined in the text. The red
  density plot gives the probability distributions that assume the NS $M$
  and $R$ data along with 68\% contour lines (solid black) and 95\%
  contour lines (dotted black). Also shown are the 68\% (blue dashed)
  and 95\% (blue dot-dashed) contour lines correspond to the
  distribution given the third data set with only a measurement of
  $I=(70{\pm}7)$ M$_{\odot}$ km$^{2}$.
\label{fig:hist2}
}
\end{figure}

As shown in Fig.~\ref{fig:hist2}, we find a strong correlation between
the radius of a 1.4-$\mathrm{M}_{\odot}$ neutron star and the pressure
at $\varepsilon=300~\mathrm{MeV}/\mathrm{fm}^{3}$ approximately twice
the nuclear saturation density (lower-left panel). There is a slightly
weaker, and more model-dependent, correlation between $\Delta I/I$ and
the transition pressure (upper-left panel), and the correlation
between $\Delta I/I$ and $L$ is extremely weak (upper-right panel), as
found in Ref.~\cite{Fattoyev10}. There is a significant range in crust
thicknesses due to the EOS, even for a fixed mass and radius, as shown
in the lower-right panel.

\section{Discussion}

There is a quandary with pulsar glitches which originates in two
results. The first is that some EOS models (such as that of 
Akmal-Pandharipande-Ravenhall~\cite{Akmal98}) have small enough crusts that the fraction of the
NS's moment of inertia contained in the crust is somewhat small
($\Delta I/I<0.05$ for a 1.4-$\mathrm{M}_{\odot}$ NS). The second is
that there is a large amount of entrainment of superfluid neutrons by
the lattice~\cite{Chamel05,Chamel12}, thus the amount of matter in the
crust which is not strongly coupled to the lattice is only 15\%$-$25\% of
the total. Together, these limit the magnitude of pulsar glitches to
be smaller than those already observed in the Vela pulsar which
requires $\Delta I/I \geq
0.016$~\cite{Link99,Espinoza11,Andersson12,Chamel13}. As can be seen
in the upper-right of Fig.~\ref{fig:hist2}, NS mass and radius data
predict a similar outcome, and the quandary stands. If we assume,
however, that systematic uncertainties invalidate current mass and
radius observations (as implied by Ref.~\cite{Suleimanov11}) and use
our third data set which only contains a measurement of the moment of
inertia of $I=(70{\pm}7)~\mathrm{M}_{\odot}~\mathrm{km}^2$, then we
find many models with $\Delta I/I > 0.09$ as also shown in the
upper-right panel. Smaller mass NSs give even larger values of $\Delta
I/I$. As with the tidal deformabilities above, assuming a measurement
of $I=(90{\pm}9)~\mathrm{M}_{\odot}~\mathrm{km}^2$ implies that
$\Delta I/I$ could be larger than 0.10. Values as large as 0.11 can be
obtained for lower mass neutron stars. These large values of $\Delta
I/I$ can accommodate the observations of glitches in Vela even with
the most extreme amounts of entrainment obtained in
Ref.~\cite{Chamel05}. A similar conclusion has also been obtained
independently in Ref.~\cite{Piekarewicz14}.

The thermal evolution of a NS crust as it cools depends on the
hydrostatic structure of the crust (as well as on how photons and
neutrons are transported). Frequently, crust cooling is studied by
using a small subset of the full variation possible for the
hydrostatic structure~\cite{Brown09,Page13}. We find that, even for a
fixed NS mass and radius, there is still considerable variation (due
to the uncertainty in the EOS of dense matter) in the thickness of the
crust. This is shown in the lower right panel of Fig.~\ref{fig:hist2}
where the probability distribution of the radius of a 1.4-$\mathrm{M}_{\odot}$
NS is plotted versus the crust thickness $\Delta R$. We find that,
for a NS with an 11 km radius, the crust thickness varies by 42\%.
This means that a more complete variation in the parameter space may
be required to determine the properties of the crust from crust
cooling observations.

If a measurement of the moment of inertia of PSR
J0737-3039A was far outside our predicted range, then that implies a
conflict with the mass and radius observations. This conflict could be
resolved with modification of strong-field GR. However, this
modification may have to be finely tuned in order to modify the NS
structure without spoiling the agreement with GR found in the
observations of the post-Keplerian parameters in the PSR J0737-3039
system~\cite{Kramer09}.

Current NS mass and radius observations are subject to several strong
systematic uncertainties (as described in
Refs.~\cite{Steiner10,Steiner13,Lattimer14b}) and a moment of inertia
measurement outside our predicted range could shed light on these
systematics. Our understanding of NS structure would be best served by
several different kinds of observations with different systematic
uncertainties so that no one effect could dominate the results. The
same reasoning given above also holds true for measurements of tidal
deformabilities, crust thicknesses, and crustal fractions of the
moment of inertia. Also, there are other neutron star observations
which we could have used, but these are unlikely to strongly modify
our results. For example, there are constraints from pulse profile
modeling on the neutron star PSR J0437-4715~\cite{Bogdanov13}, but
these are more than likely consistent with our results so long as the
mass of this particular star is near 1 M$_{\odot}$. In particular,
the 68\% limit for the radius of a 1-$\mathrm{M}_{\odot}$ star from 
that measurement is 11.3 to 14 km, and this overlaps the ranges 
given for all of the models presented in Table I. The exception to
this is if the systematic uncertainties in the qLMXB and PRE burst
observations are so large that the associated constraints on mass and
radius should be ignored {\em and} a moment of inertia measurement was
made for a lower mass star which was relatively small (i.e. $I <
70~\mathrm{M}_{\odot}~\mathrm{km}^2$ for a 1.4 $\mathrm{M}_{\odot}$
neutron star.

A large increase in the NS maximum mass, such as that implied by
Refs.~\cite{vanKerkwijk11,Romani12}, would significantly change these
results. Larger maximum masses imply larger radii (larger pressure is
needed at smaller densities to compete with gravity as the
mass becomes larger), and thus larger moments of inertia and tidal
deformabilities.

\section{Acknowledgments}

We thank N. Chamel, C. Fryer, W.C.G. Ho, J.M. Lattimer, S. Reddy and
J.R. Stone for helpful discussions. A.W.S. is supported by DOE Grant
No. DE-FG02-00ER41132. S.G. is supported by DOE Grants No.
DE-AC02-05CH11231, by the NUCLEI SciDAC program, and by the LANL LDRD
program. F.J.F. is supported by the NSF under Grant No. PHY-1068022.
F.J.F. and W.G.N. are supported by NASA through the Science Mission
Directorate under Grant No. NNX11AC41G. This research used resources
of the National Energy Research Scientific Computing Center, which is
supported by the Office of Science of the U.S. Department of Energy
under Contract No. DE-AC02-05CH11231.

\bibliographystyle{apsrev}
\bibliography{paper} 

\end{document}